\documentclass[prl,twocolumn,a4paper,aps,showpacs,superscriptaddress]{revtex4}
\usepackage[dvipdf]{graphicx}
\begin{document}

\title{Observation of soft magnetorotons in bilayer quantum Hall ferromagnets}

\author{Stefano Luin}
\affiliation{NEST-INFM and Scuola Normale Superiore, Piazza dei
Cavalieri 7, I-56126 Pisa (Italy)}
\affiliation{Bell Labs, Lucent Technologies, Murray Hill, New
Jersey 07974}
\author{Vittorio Pellegrini}
\affiliation{NEST-INFM and Scuola Normale Superiore, Piazza dei
Cavalieri 7, I-56126 Pisa (Italy)}
\author{Aron Pinczuk}
\affiliation{Bell Labs, Lucent Technologies, Murray Hill, New
Jersey 07974}
\affiliation{Dept. of Physics, Dept. of Appl. Phys.
and Appl. Math, Columbia University, New York, New York 10027}
\author{Brian S. Dennis}
\author{Loren N. Pfeiffer}
\author{Ken W. West}
\affiliation{Bell Labs, Lucent Technologies, Murray Hill, New
Jersey 07974}

\date{\today}

\begin{abstract}
Inelastic light scattering measurements of low-lying collective
excitations of electron double layers in the quantum Hall state at
total filling $\nu_T$=1 reveal a deep magnetoroton in the dispersion of
charge-density excitations across the tunneling gap. The roton
softens and sharpens markedly when the phase boundary for
transitions to highly correlated compressible states is
approached. The findings are interpreted with Hartree-Fock
evaluations that link soft magnetorotons to enhanced excitonic
Coulomb interactions and to quantum phase transitions in the
ferromagnetic bilayers.
\end{abstract}


\maketitle
%
The quantum Hall states of the two-dimensional electron gas (2DEG)
occur in high perpendicular magnetic fields that quantize the
kinetic energy into discrete, highly-degenerate Landau levels (LLs).
The energy scale for Coulomb interactions is here  $e^{2}/
\epsilon l_{B}$, where $l_B$=$\sqrt {\hbar c/eB}$ is the magnetic
length and B the perpendicular magnetic field. The neutral
quasiparticle-quasihole excitations carry the fingerprints of
electron interactions \cite{Girvinbook,persp}. Low-lying
collective modes of energies $\omega {\left({\bf q}\right)}$ and
in-plane wave vector ${\bf q}$ are linked to the condensation into
highly correlated states that emerge in the presence of strong
electron interactions. Theoretical dispersions $\omega {\left({\bf
q}\right)}$ display characteristic magnetoroton (MR) minima at
finite wave-vectors ($q \!\sim\! l_B^{-1}$) that are due to
excitonic binding terms of the Coulomb interactions in the neutral
pairs \cite{kall84, Girvin86}. It has been predicted that MRs can
soften and create instabilities leading to quantum phase
transitions that transform the ground-states into highly
correlated electron phases \cite{Girvin86, Fert89, Brey90, MacD90,
Jain97, Jogl02}.
\par
Coupled electron bilayers at total Landau level filling factor
$\nu_T$=1 exhibit a rich quantum phase diagram due to the
interplay of transition energies $\Delta_{\textrm{\tiny SAS}}$
across the tunneling gap with intra- and inter-layer interactions
\cite{MacD90, Brey90, Boebinger, Brey93, Murph94, Kun96}.
Interactions drive quantum phase transitions from the
incompressible ferromagnetic quantized Hall phase, stable at low
inter-layer spacing $d$ or large $\Delta_{\textrm{\tiny SAS}}$, to
a compressible phase that results from the collapse of the
many-body tunneling gap. In current
theories the phase transitions are linked to soft roton
instabilities in the charge-density-excitations (CDE) across the
tunneling gap \cite{MacD90, Brey90}. Within the Hartree-Fock
framework the magnetoroton instability is related to intra-layer
interactions that lead to large excitonic bindings between
quasiparticles and quasiholes.
\par
Recent experimental studies of coupled electron double layers in
$\nu_T$=1 ferromagnetic states focus on the very low
$\Delta_{\textrm{\tiny SAS}}$ region of the phase diagram, where
inter-layer Coulomb correlations are important. These studies have
displayed enhanced zero-bias inter-layer tunneling characteristics
and anomalous quantized Hall drag \cite{Spiel00, Spiel01, Kell02}.
These remarkable results are interpreted as evidence of a
Goldstone mode in the incompressible phase and of condensation of
the bilayers into many-body exciton phases.  Experiments that
probe dispersive collective excitations and their softening as
a function of $\Delta_{\textrm{\tiny SAS}}$ and $d$ could provide
direct evidence of the impact of excitonic terms of interactions
in the quantum phase transitions of the ferromagnetic bilayers.
\par
Resonant inelastic light scattering methods have been employed in
studies of very low-energy $q \sim 0$ tunneling excitations with
spin reversal in electron bilayers at even integer values of
$\nu_T$ \cite{PELLprl}. These studies have revealed that excitonic
interactions can drive changes in the quantum ground state and
finite temperature transitions \cite{Sarma97,PELLScience}. We
report here light scattering experiments that offer direct
evidence of soft magnetorotons in the CDE modes across the
tunneling gap of coupled electron bi-layers at $\nu_T$=1. MR
modes at wavevectors $q \!\sim\! l_B^{-1}$ and modes with larger
wave vectors can be accessed in the resonant light scattering
experiments due to breakdown of wave vector conservation. The
light scattering spectra with breakdown of wave vector
conservation show maxima at the critical points in the mode
dispersion \cite{marm92}.
\par
We find that the light scattering spectra of low-lying modes
typically display three bands of CDE. One is the $q \sim 0$
excitation. The other two are assigned to the CDE modes at
critical points in the dispersion. The lowest of these two is the
critical point at the magnetoroton minimum with $q \!\sim\!
l_B^{-1}$, and the higher energy band is the large density of
states of modes with $q \gg l_B^{-1}$. The MR mode softens
markedly when $\Delta_{\textrm{\tiny SAS}}$ is reduced and the
double-layer system approaches the incompressible-compressible
phase boundary. Close to this boundary the MR occurs at an
energy significantly lower than $\Delta_{\textrm{\tiny SAS}}$.
Magnetoroton spectral lineshapes and temperature dependences
display striking differences with those of the long-wavelength
modes. Close to the phase boundary the MR band shows extreme
narrowing to a width of less than 70 $\mu eV$ at 
temperatures $T\simeq 60mK$.
\par
We have interpreted these results within the framework of a
time-dependent Hartree-Fock approximation (TDHFA) that includes
breakdown of wave vector conservation in light scattering. The
calculations reproduce the MR energies and indicate that the
sharpening of the MR band reflects significant changes in the mode
dispersion and matrix element near the incompressible-compressible phase boundary.
These results uncover significant evidence that softening of the
rotons play major roles in the phase transitions of bilayers
at $\nu_T$=1 and suggest a leading role for excitonic Coulomb
interactions in transitions between highly correlated phases.
\par
\begin{figure}
\includegraphics[width=7.6cm]{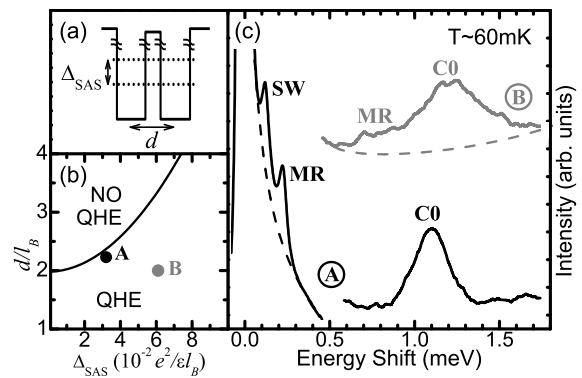}%
\caption{\label{summ} (a) Schematic representation of the double
quantum well and of the two lowest confined states. (b) Phase
diagram for the incompressible (QHE)-compressible (NO QHE) states
of the bilayer at $\nu=1$ (from Refs.~\onlinecite{Murph94,
Kun96}); the dots mark the positions of samples A and B. (c)
Inelastic light scattering spectra from sample A (black curves and
labels) and B (gray curve and labels). SW is the spin-wave, C0 is
the $q \!\sim\! 0$ CDE mode, and MR the magnetoroton minimum.}
\end{figure}

Results obtained in two modulation doped double quantum wells
(DQWs) grown by molecular beam epitaxy are presented. Samples
consist of two $18\,$nm GaAs wells separated by an undoped
Al$_{0.1}$Ga$_{0.9}$As barrier ($7.5\,$nm for sample A and
$6.23\,$nm for sample B). Figure~\ref{summ}(a) shows the schematic
profile of the bottom of the conduction band in the DQWs. Dotted
lines represent the energy of lowest symmetric and antisymmetric
states. By design the samples have the relatively high
$\Delta_{\textrm{\tiny SAS}}$ of $0.32\,$meV in sample A and
$0.58\,$meV in sample B. Magneto-transport confirms that both
samples are in the quantum Hall side of the phase diagram as shown
in Fig.~\ref{summ}(b). Total sheet densities are
$1.2\times10^{11}\,$cm$^{-2}$ in sample A and
$1.1\times10^{11}\,$cm$^{-2}$ in sample B with mobilities larger
than $10^6\,$cm$^2$/V$\,$s. Inelastic light scattering spectra are
obtained in a back-scattering geometry with light propagating
along the magnetic field. Samples are mounted in a $^3$He/$^4$He 
dilution cryo-magnetic
system with optical windows, at a small tilt angle ($20\,$degrees)
with respect to the incoming laser light. Accessible temperatures
are in the range $50\,$mK--$1.4\,$K. For these measurements the
optical emission of a dye laser is tuned to a frequency $\omega_I$
close to the fundamental interband transitions of the DQW.
Incident power densities are kept below
$\alt\!\!10^{-4}\,$W/cm$^2$, and spectra are recorded using a
double monochromator, CCD multichannel detection and spectral
resolution of $15\,\mu$eV.%
\par
\begin{figure}
\includegraphics[width=7.6cm]{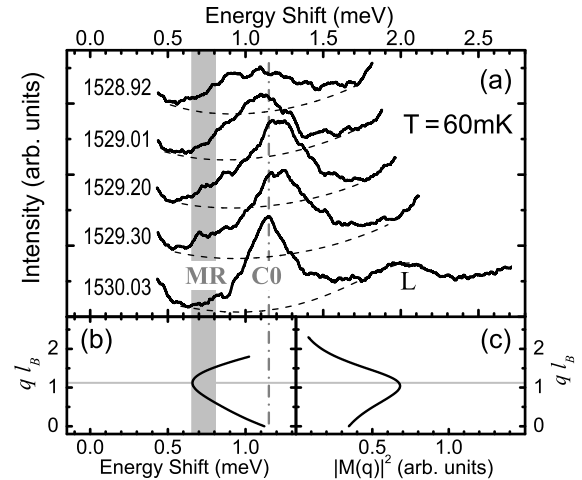}%
\caption{\label{far} (a) Inelastic light scattering spectra at
$\nu_T$=1 obtained in sample B for different incident photon
energies (in meV). The band labelled L is luminescence. (b)
Calculated dispersion of CDE modes across the tunneling gap. (c)
Calculated matrix element M(q) for inelastic light scattering. The
vertical dash-dotted line shows the energy of the $q$=0 CDE mode;
the gray area indicates contributions of large wavevector modes
with energies down to the roton minimum. The horizontal thin lines
in (b) and (c) are at the magnetoroton (MR) wave vector.}
\end{figure}

The C0 bands shown in Fig. \ref{summ}(c) have similar energies and
widths in the two samples, and also occur in spectra obtained at
B=0. On this basis they are assigned to $q\rightarrow 0$ CDE modes
\cite {Plaut97}. The structures labelled MR are remarkably
different in the two samples. They appear as a weak shoulder with
a cutoff at $0.65\,$meV in sample B, and as a sharp low-energy
peak at $0.22\,$meV in sample A. The spin wave (SW) at the Zeeman
energy $E_{Z} = 0.11\,$meV also occurs in the low-energy spectra of
sample A. Figure \ref{far}(a) shows a resonant enhancement profile
measured in sample B that reveals a characteristic outgoing
resonance with the higher optical interband transition of the
luminescence peak labelled L. The spectra from sample A display
similar outgoing resonances.
\par
Figures \ref{far}(b) and (c) show the calculated dispersion
$\omega_C\!\left(q\right)$ obtained within TDHFA and the
$\left|M\left(q\right)\right|^2$ factor that enters in the
expression of the dynamic structure factor
$S\left(q,\omega\right)$%
\begin{equation} \label{eqS}
S\left(q,\omega\right) \propto
\frac{\left|M\left(q\right)\right|^2\omega_C\!\left(q\right)\,\omega\Gamma}
{\left[\omega^2-\omega_C^2\!\left(q\right)\right]^2+\omega^2\Gamma^2}\,,
\end{equation}%
where $\Gamma$ is the homogeneous broadening \cite{Brey93,marm92}.
This TDHFA model defines a phase boundary for the instability
at values of $d/l_B$ lower than the experimental ones by a factor
of two. To correct for this discrepancy the $d/l_B$
parameters used in the calculations have been consistently
adjusted to match the parameters in our two samples. The matrix
element $\left|M\left(q\right)\right|^2$ acts as an oscillator
strength for inelastic light scattering by the collective
excitations. At the lowest order, the light-scattering cross section 
is proportional to the product of 
$S\left(q,\omega\right)$ and a factor that incorporates resonant enhancements
and optical matrix elements. $S\left(q,\omega\right)$ is used to evaluate the
intensities of inelastic light scattering by CDE modes of
different wave vectors. In this evaluation the extent of breakdown
of wavevector conservation is treated as in Ref. \cite{marm92}.
\par
The comparison of spectra of sample B in Fig.~\ref{far} with the
TDHFA calculation confirms the assignment of the band labelled C0
at $1.13\,$meV as the long wavelength CDE shifted above
$\Delta_{\textrm{\tiny SAS}}$ by dynamical many-body
contributions. This peak is observed at temperatures of up to $1.4\,$K
and over a relatively broad range of $\omega_I$ and explored
magnetic fields ($0.7 < \nu_T < 1.2$). The low-energy shoulder with
a cutoff at MR is observed only in a very narrow range of
$\omega_I$ and for B values very close to $\nu_T$=1. This
structure is assigned to resonant inelastic light scattering
processes with breakdown of wave vector conservation. The lowest
measured energy in this relatively broad structure represents the
magnetoroton in the dispersion of CDE modes.
\begin{figure}
\includegraphics[width=7.6cm]{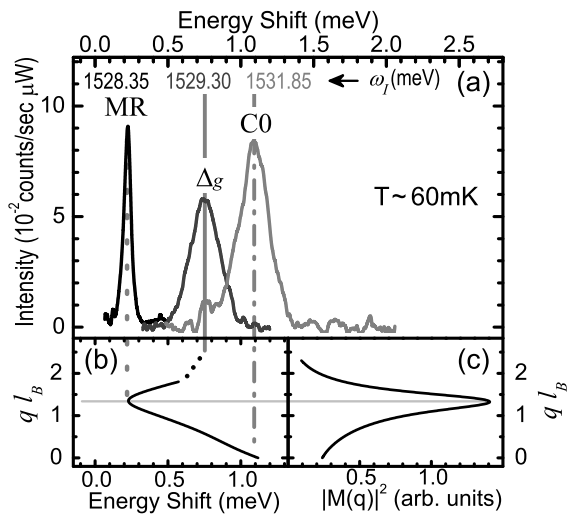}%
\caption{\label{close}(a) Light scattering bands of CDE in sample
A at $\nu_T$=1. The estimated background due to luminescence and
laser has been substracted. The incident photon energies
$\omega_I$ are indicated in meV. $\Delta_g$ labels the
$q\rightarrow\infty$ mode. (b) Calculated dispersion of CDE modes;
extrapolation to the long wave-vector limit is shown as dotted line.
(c) Calculated inelastic light scattering ``matrix element''.
Vertical lines in (a) show the peak position of CDE modes. The
horizontal line in (b) and (c) is at the MR wave vector.}
\end{figure}
\par
The marked softening of the MR mode in sample A seen in
Fig.~\ref{summ}(c) suggests links between soft magnetorotons and
the incompressible-compressible quantum phase transition at $\nu
_T=1$. Figure~\ref{close}(a) shows resonant inelastic light
spectra of CDE modes in sample A with conventional substraction of
the background due to the laser and to the main
magneto-luminescence. The results display clearly the three bands
of CDE collective modes. In addition to the CDE at $q\approx 0$
(C0, dash-dotted line at $1.08\,$meV) two lower-energy excitations
are clearly seen. The lowest energy mode at $0.22\,$meV is
assigned to the MR critical point in the dispersion. Its energy is
much lower than the MR in sample B. The MR in Fig.~\ref{close}(a)
is extremely narrow, with a full width at half maximum (FWHM) of
$\sim\!0.06\,$meV, which is a factor of three smaller than the
FWHM of the C0 band. The peak at $0.75\,$meV, labelled $\Delta_g$,
is the large wavevector CDE excitation. The $\Delta_g$ and MR
modes display a marked sensitivity on deviations of magnetic field
values from $\nu_T$=1 in a manner that is similar to the QH
states. The assignments of the C0 and MR bands in
Fig.~\ref{close}(a) are also supported by the calculated
dispersions shown in Fig.~\ref{close}(b).
\par
Significant insights are gained from a study of the q-dependence
of $\left|M\left(q\right)\right|^2$. Comparison of the calculation
for sample A in Fig.~\ref{close}(c) with that for sample B in
Fig.~\ref{far}(c) indicates that on reduction of
$\Delta_{\textrm{\tiny SAS}}$, approaching the phase
transition, $\left|M\left(q\right)\right|^2$ tends to peak sharply
at the MR wavevector. These results suggest that, by predicting the
softening and sharpening of the magnetoroton, the TDHFA, a
leading-order many body calculation, provides a framework to analyze
manifestations of interactions that eventually lead to the
soft-mode driven quantum phase transition \cite{wang2002}.
\par
The interactions that interpret the softening and narrowing of the
MR mode in sample A would eventually drive the
incompressible-compressible transition upon further reduction of
$\Delta_{\textrm{\tiny SAS}}$. These interactions are related to
the excitonic Coulomb term that creates the roton in the CDE mode
dispersion. It is conceivable that such excitonic binding increases at lower values of
$\Delta_{\textrm{\tiny SAS}}$ due to enhanced overlap between the
single-particle wavefunctions of symmetric and antisymmetric
states.
\par
The TDHFA interprets the energies, linewidths and intensities of
MRs in the light-scattering spectra. The results support the
picture that the ground state of the bilayers at $\nu_T$=1 evolves
towards a broken-symmetry state caused by the collapse of the
energy of tunneling excitations \cite{Brey90, MacD90, Brey93}. The marked
narrowing of the MR band and its interpretation within the TDHFA
suggest that the transition might be characterized by a roton wave
vector $q_{R}\sim\!{l_B}^{-1}$. Our results also support a scenario
in which the instability is associated with the
condensation of neutral excitons \cite{Spiel00, Spiel01, Kell02,
fogler01}. The exciton fluid would be linked to large densities of
extremely low energy magnetoroton modes of the incompressible
phase.
\par
\begin{figure}
\includegraphics[width=7.6cm]{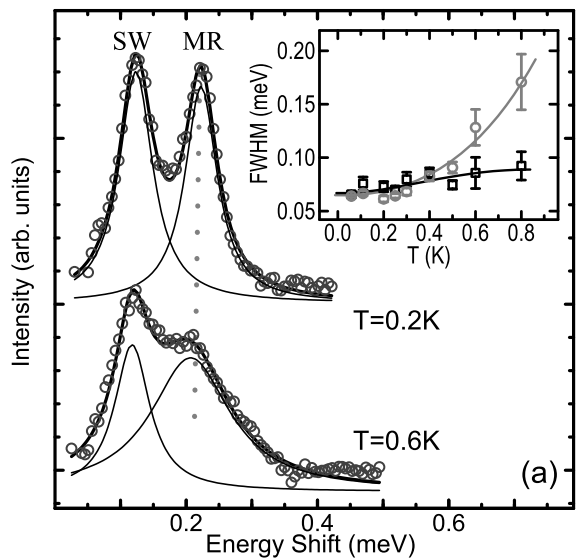}%
\caption{\label{Tdep} Open circles show the temperature dependence
of light scattering spectra in sample A. The background
substraction is as in Fig.~\ref{close}(a). The solid lines are
results of fits with two lorentzians. Inset: temperature
dependence of the FWHM for the spin-wave (SW) peak at 
the Zeeman energy(black empty squares) and the MR peak (gray empty circles). 
The solid lines are guides for the eyes and the error bars are standard deviations for
results on different measurements and with different background
subtractions.}
\end{figure}%
The MR temperature dependence measured in sample A suggests that a
highly correlated QH state may occur near the phase boundary.
Figure~\ref{Tdep} shows two representative spectra in which both
SW and MR modes are observed. The linewidth of the two excitations
have the very different temperature dependence shown in the inset
to Fig.~\ref{Tdep}. The FWHM's are obtained from Lorentzian fits
(solid lines in Fig.~\ref{Tdep}). For temperatures above 0.8K the
MR mode can no longer be observed and minor changes occur in the
SW peak. Similar temperature dependences characterize the
$q\rightarrow\infty$ ($\Delta _g$) mode and the magneto-transport
data. The characteristic temperature is here much smaller than
$\Delta _g$ and more than a factor of three below the MR energy
in sample A. The $q\sim0$ (C0) mode is temperature-independent up to
$1.4\,$K. A similar anomalous behavior observed in activated magneto-transport
was interpreted as evidence for a finite-temperature transition towards
an uncorrelated state \cite{Lay94, Abolf00}. The evolution of MR
linewidth shown in the inset of Fig.~\ref{Tdep} supports this
conclusion. From the smooth increase of the linewidth as a
function of temperature, it can be argued that thermal
fluctuations destroy the incompressible state and trigger a
continuous transition at finite temperature. Thermal excitation
of long wavelength SW modes, fixed by Larmor theorem at the Zeeman
energy, could play roles in such transition.
\par
In conclusion, we observed magnetorotons in the dispersions of
low-lying charge-density excitations in ferromagnetic electron
bilayers at $\nu_T$=1. Soft and sharp magnetorotons exist in
close proximity to the incompressible-to-compressible quantum
phase transition. The key features of the experiments have been
interpreted within time-dependent Hartree-Fock approximation.
These results suggest direct links between transitions in the
electron quantum ground-state and the low-lying dispersive
collective modes.

We are grateful to S. Das Sarma, E. Demler, S. M. Girvin,
S.H. Simon and D.W. Wang for critical reading of the
manuscript and significant suggestions.
Partial support from CNR (Consiglio Nazionale delle Ricerche),
INFM/E (Istituto Nazionale per la Fisica della Materia,
section E) and MIUR is also acknowledged.

\bibliography{PRLuin}
\end{document}